\documentclass[a4paper]{jpconf}
\usepackage{graphicx}

\begin{document}

\title{The strongly intensive observable in pp collisions at
LHC energies in the string fusion model}

\author{V V Vechernin$^1$, S N Belokurova$^1$}
\address{$^1$ Physics Faculty, Saint-Petersburg State University, 7-9 Universitetskaya emb., 199034 St.Petersburg, Russia}

\ead{v.vechernin@spbu.ru}

\begin{abstract}
The properties of the strongly intensive variable
characterizing correlations between the number
of particles in two separated rapidity interval in pp interactions
at LHC energies are studied
in the framework of the string fusion model.
We perform the MC simulations of string distributions
in the impact parameter plane to take into account
the experimental conditions of pp collisions.
We account the string fusion processes,
leading to the formation of string clusters,
embedding a finite lattice (a grid) in the impact parameter plane.
As a result, we found the dependence of this variable
both on the distance between the centers of the observation windows and their width 
for the minbias pp collisions
at several initial energies.
Analyzing
these dependencies
we can extract
the important information on the
properties
  of string clusters.
We show that in pp collisions at LHC energies the string fusion effects
have a significant impact on the behavior of this strongly intensive variable.
The role of these effects is increasing with the initial energy and centrality of collisions.
In particular, we found that the increase of this variable with initial energy
takes place due to
the growth
of the portion of the fused string clusters
in string configurations arising in pp interactions.

\end{abstract}

\def\bc{\begin{center}}
\def\ec{\end{center}}
\def\beq{\begin{equation}}
\def\eeq{\end{equation}}
\def\noi{\noindent}
\def\hs#1{\hspace*{#1cm}}

\def\av#1{\langle #1 \rangle}
\def\avL#1{{\left\langle #1 \right\rangle}}
\def\avo#1{{\av{#1}}}
\def\avr#1#2{\langle {#1} \rangle_{#2}}
\def\avup#1#2{\langle {#1} \rangle^{#2}}
\def\avC#1#2#3{\langle {#1} \rangle^{#2}_{#3}}
\def\ii{\textrm{i}}

\def\Av#1{{\left\langle #1 \right\rangle}}
\def\Avr#1#2{\left\langle {#1} \right\rangle_{#2}}
\def\Avup#1#2{\left\langle {#1} \right\rangle^{#2}}
\def\AvC#1#2#3{\left\langle {#1} \right\rangle^{#2}_{#3}}

\def\sr#1#2{{\left[{#1}\right]^{}_{#2}}}

\def\nF{{n_F^{}}}
\def\nFF{{n_F^{2}}}
\def\nB{{n_B^{}}}
\def\nBB{{n_B^{2}}}
\def\ovn{{\overline{n}}}
\def\ovp{{\overline{p}}}
\def\onB{{\overline{n}_B^{}}}
\def\onF{{\overline{n}_F^{}}}
\def\nFi{{n_i^{F}}}
\def\nBi{{n_i^{B}}}
\def\niF{{n_i^F}}
\def\niB{{n_i^B}}
\def\nkB{{n_k^B}}
\def\nkF{{n_k^F}}
\def\dniF{{d_{n_i^F}}}
\def\pijF{{p_i^{jF}}}
\def\nFis{{n_i^{\ast F}}}
\def\niFs{{n_i^{\ast F}}}
\def\etais{{\eta_i^{\ast}}}
\def\dni{{d_{n_i}}}
\def\niB{{n_i^B}}
\def\dniB{{d_{n_i^B}}}
\def\dniF{{d_{n_i^F}}}
\def\dni{{d_{n_i}}}
\def\dnk{{d_{n_k}}}
\def\oni{{\overline{n}_i}}
\def\onk{{\overline{n}_k}}
\def\oniF{{\overline{n}_i^F}}
\def\oniB{{\overline{n}_i^B}}
\def\onkF{{\overline{n}_i^F}}
\def\onkB{{\overline{n}_k^B}}
\def\ovrho{\overline{\rho}}
\def\orho{\overline{\rho}}

\def\nFr{{\frac{\nF}{\av{\nF}}}}
\def\nBr{{\frac{\nB}{\av{\nB}}}}
\def\pF{{p_{F}^{}}}
\def\ptF{{p_{tF}^{}}}
\def\pFF{{p_{F}^{2}}}
\def\pB{{p_{B}^{}}}
\def\ptB{{p_{tB}^{}}}
\def\opi{{\overline{p}_i}}
\def\ovp{{\overline{p}}}
\def\Dpi{{d_{p_i}}}
\def\dpi{{d_{p_i}}}

\def\zs{{z^\ast}}
\def\zi{{z_i}}
\def\zsi{{z^\ast_i}}
\def\ros{{\rho^\ast}}
\def\roi{{\rho_i}}
\def\rosi{{\rho^\ast_i}}
\def\rhoiF{{\rho^F_i}}
\def\rhoiB{{\rho^B_i}}
\def\rhokF{{\rho^F_k}}
\def\rhokB{{\rho^B_k}}

\def\sumetai{{\sum_{\eta_i=1}^{\infty}}}
\def\sumetak{{\sum_{\eta_k=1}^{\infty}}}

\def\sumiM{{\sum_{i=1}^M}}
\def\sumkM{{\sum_{k=1}^M}}
\def\sumi{{\sum_{i}^{}}}
\def\sumk{{\sum_{k}^{}}}
\def\sumink{{\sum_{i\neq k}^{}}}
\def\sumik{{\sum_{i, k}^{}}}
\def\sumj{{\sum_{j=1}^{n_i}}}
\def\sumjB{{\sum_{j=1}^{n^B_i}}}
\def\sumjF{{\sum_{j=1}^{n^F_i}}}
\def\sumjj{{\sum_{j'=1}^{n_k}}}
\def\sumjnj{{\sum_{j\neq j'}^{n_i}}}

\def\SS#1#2{{S_{#1/#2}}}
\def\SSM#1#2{{S_{-#1/#2}}}
\def\SSS#1#2{{S^{\,2}_{#1/#2}}}

\def\deta{{\delta\eta}}
\def\etai{{\eta_i}}
\def\oeta{{\overline{\eta}}}
\def\oetai{{\overline{\eta}_i}}
\def\oetak{{\overline{\eta}_k}}
\def\detai{{d_{{\eta}_i}}}
\def\Deta{{\Delta\eta}}

\def\Ceta{{C_{\eta}}}
\def\Cetas{{C_{\eta^{\ast}}}}
\def\Cetab{{C_{\overline{\eta}}}}
\def\CnF{{C^{F}_{n}}}
\def\Cn{{C_n}}
\def\CnFs{{C^{F}_{n^\ast}}}
\def\CnB{{C^{B}_{n}}}
\def\CpF{{C^{F}_{p}}}
\def\CpB{{C^{B}_{p}}}
\def\ConB{{C_{\overline{n}}^B}}
\def\ConF{{C_{\overline{n}}^F}}

\def\pir{{\frac{1}{2\pi}}}
\def\pirr{{(2\pi)^{-2}_{}}}

\def\dpF{{\delta p_{{\rm T}F}^{}}}
\def\dpB{{\delta p_{{\rm T}B}^{}}}
\def\bpF{{\textbf{p}_{F}^{}}}
\def\bpB{{\textbf{p}_{B}^{}}}

\def\muF{{\mu^{}_F}}
\def\muB{{\mu^{}_B}}
\def\muFF{{\mu^{2}_F}}
\def\muBB{{\mu^{2}_B}}
\def\mFr{{\frac{\mF}{\av{\mF}}}}
\def\mBr{{\frac{\mB}{\av{\mB}}}}

\def\omu{{\av{\mu}}}
\def\omF{{\av{\mu^{}_F}}}
\def\omB{{\av{\mu^{}_B}}}
\def\omBF{{\av{\mu^{}_F\mu^{}_B}}}
\def\omFB{{\av{\mu^{}_F\mu^{}_B}}}

\def\mo{{\mu^{}_{0}}}
\def\moo{{\mu^{2}_{0}}}
\def\rhoo{{\rho^{}_{0}}}
\def\rhooo{{\rho^{2}_{0}}}
\def\moF{{\mu^{}_{0F}}}
\def\moB{{\mu^{}_{0B}}}

\def\ommF{{\av{\mu_F^2}}}
\def\ommB{{\av{\mu_B^2}}}
\def\omFF{{\av{\mu_F^{}}_{}^2}}
\def\omBB{{\av{\mu_B^{}}_{}^2}}
\def\df{\delta_{F,\sum F_i}}
\def\db{\delta_{B,\sum B_i}}
\def\pp{\prod_{i=1}^N p(B_i,F_i)}
\def\sumn{\sum^n_{i=1}}

\def\oN{\overline{N}}
\def\obnn{\overline{b}_{nn}}
\def\obr{\overline{b}^{rel}_{}}

\def\ob{\widetilde{b}}
\def\obnn{\widetilde{b}_{nn}}
\def\obpn{\widetilde{b}_{p_tn}}
\def\obpp{\widetilde{b}_{p_tp_t}}
\def\bnn{b_{nn}}
\def\bnnh{b^{hom}_{nn}}
\def\bpn{b_{p_tn}}
\def\bpnh{b^{hom}_{p_tn}}
\def\bpp{b_{p_tp_t}}
\def\bpph{b^{hom}_{p_tp_t}}

\def\bcor{b_{corr}}
\def\bcorr{b_{corr}^{}}
\def\bcorr{b_{corr}^{}}
\def\brel{b^{rel}}
\def\brelLR{b_{rel}^{LR}}
\def\bm{\beta_{mod}^{}}
\def\bmp{\beta_{mod}^{'}}
\def\babs{b_{abs}^{}}
\def\ba{b_{abs}^{}}
\def\bsym{b_{sym}^{}}
\def\brob{\beta_{rob}^{}}
\def\oba{\overline{b}^{abs}_{}}
\def\ar{a^{rel}_{}}
\def\aa{a^{abs}_{}}
\def\dnF{\nF-\av{\nF}}
\def\pc{\!+\!}
\def\mc{\!-\!}
\def\ppc{\!+\!...\!+\!}
\def\yFB{{\eta^{}_{FB}}}
\def\yBF{{\eta^{}_{FB}}}
\def\yF{{\eta^{}_F}}
\def\yB{{\eta^{}_B}}
\def\fFB{{\phi^{}_{FB}}}
\def\fBF{{\phi^{}_{FB}}}
\def\fF{{\phi^{}_F}}
\def\fB{{\phi^{}_B}}

\def\dyf{{d\eta d\phi}}
\def\dyfo{{d\eta_1 d\phi_1}}
\def\dyft{{d\eta_2 d\phi_2}}
\def\dyp{{d\eta'}}
\def\dfp{{d\phi'}}
\def\dyfp{{d\eta' d\phi'}}

\def\Dyy{{(\delta \eta)^2}}
\def\Dy{{\Delta y}}
\def\dy{{\delta y}}
\def\Df{{\delta \phi}}

\def\DyF{{\delta y^{}_F}}
\def\DyFF{{(\delta \eta^{}_F)^2}}
\def\DyB{{\delta y^{}_B}}
\def\ygap{{y_{gap}}}

\def\DfFF{{(\delta \phi^{}_F)^2}}
\def\DfF{{\delta \phi^{}_F}}
\def\DfB{{\delta \phi^{}_B}}

\def\dfF{{\delta \phi^{}_F}}
\def\dfB{{\delta \phi^{}_B}}

\def\dyF{{\delta y^{}_F}}
\def\dyFF{{\delta y^{2}_F}}
\def\dyB{{\delta y^{}_B}}
\def\dyBB{{\delta y^{2}_B}}

\def\DyfFF{{(\delta \eta^{}_F\delta \phi^{}_F)^2}}
\def\DyfF{{\delta \eta^{}_F\delta \phi^{}_F}}
\def\DyfB{{\delta \eta^{}_B\delta \phi^{}_B}}

\def\aFF{{\delta^{2}_F}}
\def\aFFr{{\delta^{-2}_F}}
\def\aF{{\delta^{}_F}}
\def\aB{{\delta^{}_B}}
\def\aFr{{\delta^{-1}_F}}
\def\aBr{{\delta^{-1}_B}}
\def\aW{{\delta^{}_W}}

\def\acF{{\delta \eta^{}_F\delta \phi^{}_F/2\pi}}
\def\acB{{\delta \eta^{}_B\delta \phi^{}_B/2\pi}}
\def\acW{{\delta \eta\delta \phi/2\pi}}

\def\eg{{\eta^{}_{gap}}}
\def\yg{{\eta^{}_{gap}}}
\def\fg{{\phi^{}_{gap}}}

\def\omn{{\omega_n}}
\def\omN{{\omega_N}}
\def\ometa{{\omega_\eta}}
\def\ommu{{\omega_\mu}}
\def\ommueta{{\omega^{(\eta)}_{\mu}}}
\def\ommuone{{\omega^{(1)}_{\mu}}}

\def\bp{\textbf{p}}
\def\oq{\overline{q}}
\def\fv{\phi}
\def\loy{\lambda_1(\eta)}
\def\loyo{\lambda_1(\eta_1)}
\def\loyt{\lambda_1(\eta_2)}
\def\lty{\lambda_2(\eta_1,\eta_2)}
\def\lo#1{\lambda_1(#1)}
\def\loo#1{\lambda^2_1(#1)}
\def\lt#1{\lambda_2(#1)}
\def\Lam#1{\Lambda(#1)}
\def\tLam#1{\widetilde{\Lambda}(#1)}

\def\loyf{\lambda_1(\eta,\phi)}
\def\loyfo{\lambda_1(\eta_1,\phi_1)}
\def\loyft{\lambda_1(\eta_2,\phi_2)}
\def\ltyf{\lambda_2(\eta_1,\phi_1;\eta_2,\phi_2)}

\def\roy{\rho_1(\eta)}
\def\royo{\rho_1(\eta_1)}
\def\royt{\rho_1(\eta_2)}
\def\rty{\rho_2(\eta_1,\eta_2)}
\def\ro#1{\rho_1(#1)}
\def\roo#1{\rho^2_1(#1)}
\def\rt#1{\rho_2(#1)}
\def\royf{\rho_1(\eta,\phi)}
\def\royfo{\rho_1(\eta_1,\phi_1)}
\def\royft{\rho_1(\eta_2,\phi_2)}
\def\rtyf{\rho_2(\eta_1,\phi_1;\eta_2,\phi_2)}

\def\IFF{{I_{F\!F}^{}}}
\def\IBF{{I_{F\!B}^{}}}
\def\IBB{{I_{B\!B}^{}}}
\def\JFF{{J_{F\!F}^{}}}
\def\JBF{{J_{F\!B}^{}}}
\def\JFB{{J_{F\!B}^{}}}

\def\cov{{\textrm{cov}}}

\def\SigFB{{\Sigma(\nF,\nB)}}
\def\SigN{{\Sigma_{\eta_1} (n_1^F, n_1^B)}}
\def\SigNi{{\Sigma_{\eta_i} (\niF, \niB)}}
\def\Sigmu{{\Sigma (\muF, \muB)}}
\def\Sigetamu{{\Sigma_\eta (\muF, \muB)}}
\def\Sigmuone{{\Sigma_1 (\muF, \muB)}}

\def\nFp{{n_F^+}}
\def\nFm{{n_F^-}}
\def\nBp{{n_B^+}}
\def\nBm{{n_B^-}}

\def\Sigpp{{\Sigma (n^+_F, n^+_B)}}
\def\Sigpm{{\Sigma (n^+_F, n^-_B)}}
\def\Sigmp{{\Sigma (n^-_F, n^+_B)}}
\def\Sigmm{{\Sigma (n^-_F, n^-_B)}}

\def\Spp#1{{\Sigma_{#1} (\mu^+_F, \mu^+_B)}}
\def\Spm#1{{\Sigma_{#1} (\mu^+_F, \mu^-_B)}}
\def\Smp#1{{\Sigma_{#1} (\mu^-_F, \mu^+_B)}}
\def\Smm#1{{\Sigma_{#1} (\mu^-_F, \mu^-_B)}}

\def\LamNi{{\Lambda_{\eta_i}}}
\def\lamNi{{\lambda^{\eta_i}}}
\def\lamNit{{\lambda_2^{\eta_i}}}
\def\lamNiz{{\lambda_0^{\eta_i}}}

\def\Lameta{{\Lambda_\eta}}
\def\lameta{{\lambda_\eta}}
\def\lametat{{\lambda_2^\eta}}

\def\detaF{\delta\eta_{F}}
\def\detaB{\delta\eta_{B}}
\def\JFBeta{{J_{F\!B}^{\eta}}}
\def\JFFeta{{J_{F\!F}^{\eta}}}
\def\muetaz{{\mu^\eta_0}}
\def\muonez{{\mu^1_0}}

\def\Lametaz{{\Lambda^\eta_0}}
\def\yetac{{y^\eta_{corr}}}
\def\yc{{y^1_{corr}}}
\def\ometamu{{\omega^\eta_\mu}}

\def\etac{\eta^{}_{corr}}
\def\etack{\eta^{(k)}_{corr}}
\def\etaco{\eta^{(1)}_{corr}}

\def\SigFB{{\Sigma(\nF,\nB)}}
\def\SigN{{\Sigma_{N_1} (n_1^F, n_1^B)}}
\def\SigNi{{\Sigma_{N_i} (\niF, \niB)}}
\def\Sigmu{{\Sigma (\muF, \muB)}}

\def\Sigk{{\Sigma_{k} (\muF, \muB)}}
\def\Sigmuk{{\Sigma_{k} (\muF, \muB)}}
\def\Lamk{{\Lambda_{k}}}
\def\Lamkk{{\Lambda_{0}^{(k)}}}
\def\etakc{{\eta^{(k)}_{corr}}}
\def\muk{{\mu_{0}^{(k)}}}
\def\ommuk{{\omega_\mu^{(k)}}}

\def\LamNi{{\Lambda_{N_i}}}
\def\lamNi{{\lambda^{N_i}}}
\def\lamNit{{\lambda_2^{N_i}}}
\def\lamNiz{{\lambda_0^{N_i}}}


\section{Introduction}
By now, the string (color flux tubes) model \cite{Kaidalov82,Capella94} has become a standard approach
for the description of the soft part of hadronic interactions at high energies.
Different versions of the string model are applied in the various
MC event generators: PYTHIA, VENUS, HIJNG, AMPT, EPOS etc.,
for a description of soft processes in strong interactions,
when the perturbative QCD approach does not works.

Along with this, It was also recognized that
at a large string density
in nucleus-nucleus collisions
the strings start overlap in the impact parameter plain
due to finite transverse area of a string, considered as a color flux tube.
So one has to take into account the interaction between strings
leading to the string fusion processes and the formation of string clusters (``color ropes'')
\cite{Biro84,BP93}.
In the present paper we'll argue that the string fusion processes
play important role also in pp interactions at LHC energies.

As
one of the tools
for the investigation of these string fusion effects
the study of the forward-backward (FB) correlations
between observables in two separated rapidity intervals was
suggested in \cite{PRL94}.
Really, it is known that the investigations of the long-range rapidity correlations
enable to obtain the information on
the very initial stages of a hadronic collision \cite{Dumitru08},
and, in particular, on the
emerging
string configuration.

Unfortunately, back in the work \cite{CapKr78} it was shown that
the traditional FB correlation coefficient  between the charge particle multiplicities,
$\nF$ and $\nB$, in the observation rapidity windows
strongly depends on
the event-by-event
variance
of the number of cut pomerons (strings)
in pp collisions, i.e. on
the so-called ``volume'' fluctuation
- the trivial fluctuation in the number of sources.
From this point of view it is desirable to
search
for another observables, which is not sensitive to the fluctuation
in the number of sources (strings), but is sensitive to the fluctuation of the properties of sources,
e.g. to the formation of string clusters by string fusion processes.

We can suppress the influence of these trivial ``volume'' fluctuations
going to the studies of the FB correlation coefficient  between the
intensive quantities, such as e.g. event-mean transverse momenta $\pF$ and $\pB$
of charge particles in the observation rapidity intervals $\dy_F$ and $\dy_B$
instead of $\nF$ and $\nB$, as in \cite{Q16,ICPPA16Kov},
or
defining
the more sophisticated correlation observable between $\nF$ and $\nB$.
In the present paper we'll focus on the last approach.

\section{The strongly intensive observable $\SigFB$}

The general methods of the constructing of variables not affected by the ``volume'' fluctuations
(the so-called strongly intensive observables)
were developed in the paper \cite{GorGaz11}. Later \cite{Andronov15} it was suggested to study
the strongly intensive quantity $\SigFB$, defined
for the charge particle multiplicities, $\nF$ and $\nB$, in two observation windows:
\beq
\label{SigmaFB}
\Sigma(\nF,\nB)\equiv[\av\nF\,\omega_\nB+\av\nB\,\omega_\nF-2\, \cov(\nF\,\nB)]/[\av\nF+\av\nB] \ ,
\eeq
where the $\omega_n\equiv D_n/\av n$ is a scaled variance and $\cov(\nF\,\nB)\equiv \av{\nF\nB}-\av\nF\av\nB$.
For symmetric reaction and symmetric observation windows, $\dy_F$=$\dy_B$=$\dy$, we have
$\av\nF=\av\nB\equiv \av n $,
$\omega_\nF=\omega_\nB\equiv \omega_n$
and  the definition (\ref{SigmaFB}) can be simplified to
\beq
\label{Sigma-s}
\SigFB=\omn - \cov(\nF,\nB)/\av n= [\av  {n^2}-\av{\nF\nB}]/{\av n}  \ .
\eeq

In paper \cite{Q18} it was demonstrated that in the model with independent identical strings
the observable $\SigFB$ can be expressed only through the parameters of a single string -
the one- and two-particle rapidity distributions
arising from decays of a string:
\begin{equation}
\label{lam12tr}
\lambda (y)\equiv{dN}/{dy}=\av {\mu}/\dy=\mu_0  \ ,
\hs {0.5}
\lambda_2 (y_1, y_2)\equiv{d^2N}/{dy_1\,dy_2}
=\lambda_2 (y_1\!-\!y_2)   \ .
\end{equation}
Instead of the $\lambda_2 (y_1, y_2)$ one usually exploits the two-particle correlation function of a string:
\begin{equation}
\label{Lambda}
\Lambda(\eta_1,\eta_2)\equiv{\lambda_2 (\eta_1,\eta_2)}/{\lambda(\eta_1) \lambda(\eta_2)}-1
={\lambda_2 (y_1\!-\!y_2)}/{\mu_0^2}-1 = \Lambda(\eta_1-\eta_2)
 \ ,
\end{equation}
The last transitions in these formulas corresponds to the rapidity invariant approximation,
which is fulfilled at mid-rapidities at LHC energies.
It was shown in \cite{Q18} that in the model with independent identical strings
\beq\label{Signmu}
\SigFB=\Sigmu \ .
\eeq
Here the $\Sigmu$ is the strongly intensive variable $\Sigma$
defined similarly (\ref{SigmaFB}), but for a single source: the $\muF$ and $\muB$
are the number of particles produced
in forward and backward rapidity windows
from decays of a single string. It's
connected with
the string parameters
as follows
\beq\label{Sigmu}
\Sigmu=1+\mu_0\, \dy\,  [J_{FF}-J_{FB}] =\Sigma(\dy, \Dy) \ ,
\eeq
where
\beq\label{JFBtr}
\JFB= \frac{1}{\dyF\dyB}\int_{\dyF}\!\!\!\!\!\!dy_1 \int_{\dyB} \!\!\!\!\!\!dy_2\
\Lambda(y_1\!-\!y_2)
   \ ,
\hs 1
\JFF=  \frac{1}{\dyFF}\int_{\dyF}\!\!\!\!\!\!dy_1 \int_{\dyF} \!\!\!\!\!\!dy_2\
\Lambda(y_1\!-\!y_2)
   \ .
\eeq
For symmetric case
it depends on the acceptance of the observation windows, $\dy$=$\dy_F$=$\dy_B$, and
the rapidity distance, $\Dy$, between the centers of these windows,
so
$\Dy\geq\dy$.

The formula (\ref{Signmu}) proves
that in the framework of the model with independent identical strings
the variable $\SigFB$
demonstrates
the strongly intensive property.
Its value is equal to the value for a single string $\Sigmu$ and does not depend on the number of strings, produced in a given event
and their event-by-event fluctuations.

Unfortunately, the string parameters, $\mu_0$ and $\Lambda(\eta_1-\eta_2)$, extracted
in papers \cite{NPA15,Q18,EPJA19,UNIV19}
from the ALICE data \cite{ALICE15}  on the FB correlations in pp collisions at initial energies 0.9-7 TeV
prove to be dependent on the energy.
By (\ref{Signmu})--(\ref{JFBtr}) this leads to the dependence of the $\SigFB$ on the collision energy,
what  can be interpreted as a signature of string fusion processes in pp collisions
at LHC energies \cite{Q18,EPJA19,UNIV19,TMF19}.

Really, in the paper \cite{TMF19} it was shown, that in the model with
different types of strings the expression (\ref{Signmu}) for $\SigFB$ is replaced by
\beq\label{Sigfus}
\SigFB
= \sum_{\eta=1}^{\infty} \alpha_\eta \, \Sigetamu \ ,
\eeq
where
$\Sigetamu$
is a $\Sigma$ variable for a string cluster,
arising due to
the fusion of the $\eta$ initial strings.
For each such cluster the formulas (\ref{Sigmu}) and (\ref{JFBtr}) are valid, but
with the replacement of the single string parameters to the parameters
of the corresponding cluster formed by the $\eta$ fused initial strings:
\beq\label{parfus}
\mu_0\to\mu^\eta_0 \ , \hs 1 \Lambda (y_1 - y_2)\to\Lambda_\eta (y_1 - y_2)\ .
\eeq
The
weighting factors
$\alpha_\eta=\avr n \eta/\av n$ in (\ref{Sigfus})
is an average portion of particles produced  from decays of all clusters of a given $\eta$ type.
Note that the first term with $\eta=1$ corresponds to the contribution from single strings.

Obviously, with a change in the initial energy and centrality of pp collisions
the portion of the clusters, corresponding to the different number of fused strings,
will changes, leading to the change of the weights $\alpha_\eta$
and hence by (\ref{Sigfus}) to the change of the observable $\SigFB$.
So, given the possibility 
of the formation of sources of different types
the observable $\SigFB$ becomes equal to the weighted average 
of the ones for different string clusters with weights, $\alpha_\eta$, 
which depend 
on the details of the collision - the initial energy and centrality of the collision.


\section{Modeling string configurations in pp collision}

For the Monte-Carlo generation of the string configurations arising in pp collision
we have used the approach developed earlier in the paper \cite{Dubna12Lak}.
In this approach we suppose that at given value of the impact parameter $b$
for the number of cut pomerons, $N$,
we have event-by-event the poissonian distribution with the parameter
$\overline{N} (b)=N_0\, \exp{(-b^2/2 r_0^2)}$,
discarding the cases with $N=0$. The last condition reflects the fact that we take into account
only inelastic (non-diffractive, ND) pp collisions.

We will consider that each cut pomeron corresponds to the formation
of two initial strings \cite{Kaidalov82, Capella94}, $N_{str}=2N$.
To take into account later the string fusion effects
by the MC calculations
we need to simulate not only the total number of strings in each event,
but also their distribution in the transverse plane.
According to \cite{Dubna12Lak}, we suppose that these strings are distributed in the transverse plane with the
probability density:
\begin{equation}
\label{strind density}
w_{str}({\bf s}, {\bf b}) \sim
T({\bf s}-{\bf b}/2)T({\bf s}+{\bf b}/2)/{\sigma_{\!pp}(b)} \ ,
\end{equation}
where the $w_{str}({\bf s}, {\bf b})$ is the probability density to find a string at the point ${\bf s}$ of the transverse plane
in the case of the ND pp collision at the impact parameter $b$,
the $\sigma_{\!pp}(b)$ is the probability of such interaction.
Here the $T({\bf s})$  is the partonic profile function of a nucleon, for which
we use the simplest gaussian distribution with some parameter $r^{}_0$.

As it was shown in \cite{Dubna12Lak}, in this model along with MC simulations
we can analytically calculate all quantities of physical interest.
So, for the cross-section of the ND $pp$ interaction
we have
\begin{equation}
\label{ND cross section}
\sigma^{\!N\!D}  = \int \sigma_{\!pp}^{}(b)\ d^{2}\vec{b}
= 2\pi  r_0^2 \left[  \Phi( N_{0}) + \gamma + \ln{N_0}\right] \ ,
\hs1 \Phi(x) = \int_{x}^{\infty}e^{-t}dt/t \ ,
\end{equation}
where the $\gamma=0.577...$ is the Euler constant.
For the average number of cut pomerons, $\av{N}$,
the scaled variance of the number of cut pomerons, $\omN\equiv D_N/\av{N}$,
and the probability, $P(N)$, to have $N$ cut pomerons in a non-diffractive pp collision
at some given initial energy we have, respectively:
\begin{equation}
\label{omNpom}
\av N  = { N_0}/[{\Phi( N_{0}) + \gamma + \ln{N_0}}] \ ,
\hs1 \omN= 1+{N_0}/{2} - \av  N \ ,
\end{equation}
\begin{equation}
\label{P pom}
P(N) =
\frac{2\pi r_0^2}{\sigma^{\!N\!D}_{\!pp} N}
\left[1- e^{-N_0} \sum_{l=0}^{N-1} N_0^l/l! \right]  \ .
\end{equation}

In the paper \cite{Dubna12Lak} it was also shown that
the present approach
leads to
the same distribution (\ref{P pom}) for the number of cut pomerons in minbias nondifractive pp collisions
as the Gribov-Regge approach (see e.g. \cite{Kaid-Cap-Shab}).
This enables to connect  the model parameters $N_0$ and $ r^{}_0$,
with the parameters of the pomeron trajectory and its couplings to hadrons:
\begin{equation}
\label{paramfix}
 r^{}_0=\sqrt{{2\lambda}/{C}},\
N_0=\frac{2\gamma_{pp} C}{\lambda}\exp(\xi\Delta)  ,\
\lambda=R_{pp}^2+\alpha'\xi,\ \xi=\ln (s/s^{}_0)\ .
\end{equation}
Here  $\Delta$ and $\alpha'$ are the intercept and the slope of the
pomeron trajectory.
The parameters $\gamma_{pp}$ and $R_{pp}$ characterize the coupling of the pomeron trajectory
to the initial hadrons. The quasi-eikonal parameter $C$ is related to
the small-mass diffraction dissociation of incoming hadrons, $s^{}_0\simeq 1 GeV^2$.

The numerical values of the Regge parameters in the formulae (\ref{paramfix}),
obtaining by the comparison with experimental data,
evolve with encrease of the available initial pp collision energy.
The typical modern values are as follows (see e.g. \cite{Kaid-Cap-Shab}):
\begin{equation}
\label{paramKCS}
\Delta= 0.139,\ \ \ \alpha'=0.21\ GeV^{-2},\ \ \ \gamma_{pp}=1.77\ GeV^{-2},\
\ \  R_{pp}^2=3.18\ GeV^{-2},\ \ \ C=1.5\ .
\end{equation}
However the test calculations with this set of parameters
demonstrated that we can not explaine the rapid increase of the multiplisity at mid rapidities,
experimentally observed at initial energies higher than 900 GeV.
This corresponds with the results obtained in the paper \cite{Zabrodin16},
in which for the same reasons the increased value of the
$\Delta= 0.1855$ and the reduced value of the $\alpha'=0.175\ GeV^{-2}$
for the soft pomeron were used.

\begin{table}
\caption{\label{NDn}The non-diffractive cross section, the multiplicity density at mid-rapidity and
the mean number of initial strings in $pp$ collisions at different initial energies.}
\begin{center}
\begin{tabular}{lllll}
\br
$\sqrt{s} (GeV) $ & $\sigma^{ND}_{th} (mb) $& $\sigma^{ND}_{MC} (mb) $& $dN^{ND}/dy$ & $\av{N_{str}}$  \\
\mr
 60    & 24.9 & 24.9 & 2.44 & 4.2 \\
900   & 39.9 & 39.9 & 3.76 & 7.8 \\
7000  & 52.5 & 52.4 & 5.44 & 13.4 \\
13000 & 56.5 & 56.6 & 6.03 & 16.0 \\
\br
\end{tabular}
\end{center}
\end{table}

However, within the framework of our model,
these changes of the paremeters are insufficient,
since we have additional losses in multiplicity
due to the increase of string fusion effects
with increasing energy.
So we used in our calculations the following set of the parameters:
\begin{equation}
\label{ourparam}
\Delta= 0.2,\ \ \ \alpha'=0.05\ GeV^{-2},\ \ \ \gamma_{pp}=1.035\ GeV^{-2},\
\ \  R_{pp}^2=3.3\ GeV^{-2},\ \ \ C=1.5
   \  .
\end{equation}
which gives an appropriate description
the non-diffractive cross section and the multiplicity density at mid-rapidities
in $pp$ collisions (see the Table \ref{NDn}).
The problem with the selection of these parameters
is that it is difficult to obtain a rapid increase in multiplicity with increasing energy,
while maintaining a moderate increase in the nondiffractive cross section.

Note that the mean multiplicities in the Table \ref{NDn}
are calculated already with taking into account the string fusion effects,
as it is explained in the next section with $\mu^1_0=\mu_0=0.7$.
To check the MC algorithm we have calculated the non-diffractive cross section
both analytically by formula (\ref{ND cross section}),  $\sigma^{ND}_{th}$,
and by MC simulations as $\sigma^{ND}_{MC}=S_b\, {n_{sim}(N=0)}/{n_{sim}(N\geq0)}$,
where the $n_{sim}(N=0)$ is a number of simulation without cut pomerons,
the $n_{sim}(N\geq0)$ is a total number of simulation and $S_b$ is an area
of the impact parameter simulations.

\section{Accounting the string fusion effects}

In our MC model to account the string fusion effects
we introduce
a finite transverse lattice (a grid) \cite{Vest2,EPJC04} with the cells,
which area is equal to the transverse area of a single initial string, $\sigma_{str}=\pi r^2_{str}$.
We use $r_{str} = 0.2\ fm$ in our calculation \cite{Koval-Rstr}.

Generating the event we at first generate the impact parameter $b$,
then the number of pomerons, $N$,
and hence the number of strings, $N_{str}=2\,N$,
as it was described in the previous section.
After that we distribute these strings in the transverse plane according
to the density given by the formula (\ref{strind density}).
We consider that all $\etai$ strings, which centers occur within the same $i$-th cell,
are fused into one string cluster.

In the accordance with the string fusion model prescriptions \cite{BP93}
we suppose
 the following dependence of the average number of particles, $\avr \mu \eta$, produced
 from the decay of this string cluster on the number of fused strings $\eta$:
 	\begin{equation}
		\label{mean mult}
	\avr \mu \eta=\mu^\eta_0 \delta y=\mu_0 \sqrt{\eta} \, \delta y \ ,
	\end{equation}
where the $\mu^\eta_0$ and the $\mu^1_0\equiv \mu_0$ is the average rapidity particle densities produced
from the hadronization of the string cluster and one single string.

Using generated string configurations we can find the weights
 $\alpha_\eta$ in (\ref{Sigfus}) as follows
 \beq\label{alfaMC}
\alpha_\eta
= \frac{\avr n \eta}{\sum_{\eta=1}^{\infty} \avr n \eta}
= \frac{\avr m \eta  \mu_0^\eta \dy}{\sum_{\eta=1}^{\infty} \avr m \eta  \mu_0^\eta \dy}
= \frac{\avr m \eta \sqrt{\eta}}{\sum_{\eta=1}^{\infty} \avr m \eta \sqrt{\eta}}
  \ ,
\eeq
where the $\avr m \eta$ is the mean number of clusters with $\eta$ fused strings,
which we take from our MC simulations of the string configurations.


To calculate now the $\SigFB$ by formula (\ref{Sigfus}) we have to make some
assumptions on the two-particle correlation function $\Lambda_\eta (y_1-y_2)$ of a cluster.
As in paper \cite{Q18} we'll use the simplest exponential form
\beq \label{Lameta_exp}
\Lameta(y_1-y_2)=\Lametaz \exp(-{|y_1-y_2|}/{\yetac})  \ ,
\eeq
where $\yetac$ is a characteristic correlation length.
In this case the integrals (\ref{JFBtr}) easily can be calculated analytically
and we can find all $\Sigetamu$ by formulas similar to (\ref{Sigmu}).
\begin{figure}
	\centering
	\begin{tabular}{cc}
		\hspace{-5mm}	\includegraphics[width=0.55\linewidth]{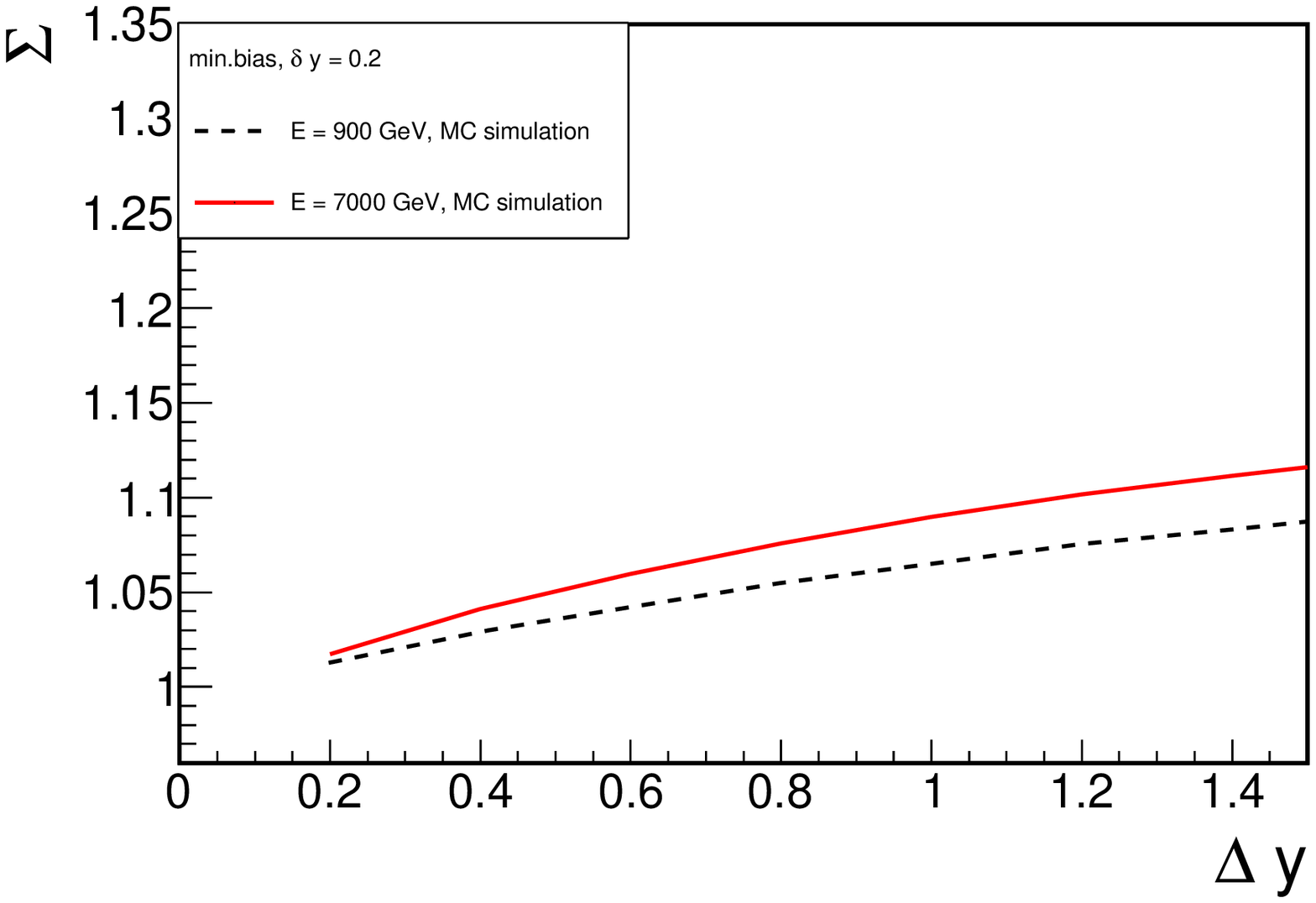}
		&	\hspace{-14mm}	\includegraphics[width=0.55\linewidth]{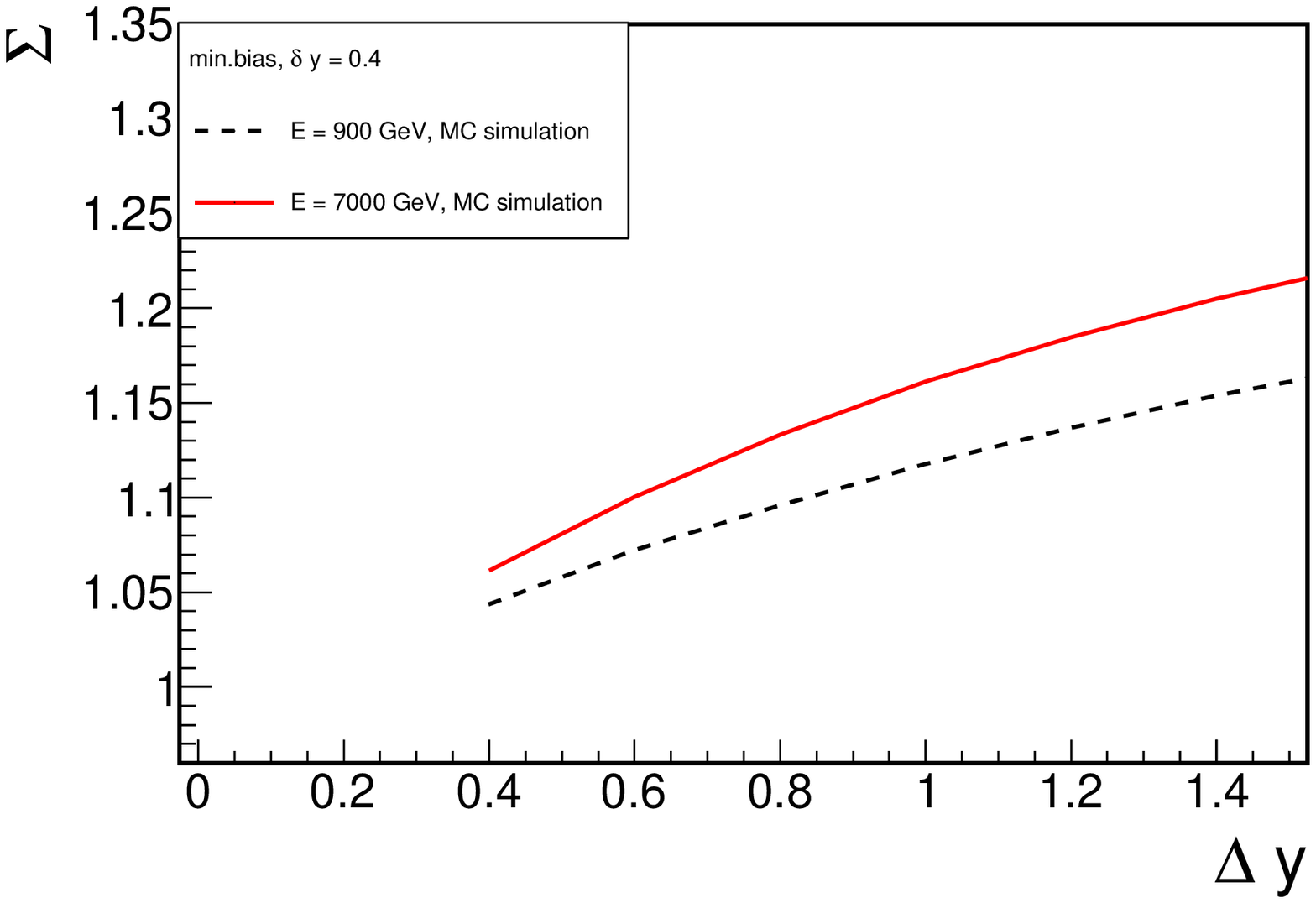}\\
	\end{tabular}
\caption{\label{Sig}
The strongly intensive observable $\SigFB$ (\ref{SigmaFB}) for pp collisions
as a function of the rapidity distance $\Dy$ between the centers of the FB observation windows,
for two widths of windows: $\dy$=0.2 (left panel) and $\dy$=0.4 (right panel),
and for two initial energies: 0.9 TeV (dashed lines) and 7 TeV (solid lines),
calculated for particles with transverse momenta
in the interval 0.3-1.5 GeV/c, as in the experimental analysis in \cite{ALICE15}.
}
\end{figure}

Since by (\ref{mean mult}) the multiplicity density from a string cluster decay
is considered proportional to $\sqrt{\eta}$
and the correlations take place only between the neighbor string cluster segments,
then we'll suppose that the characteristic correlation length $\yetac$
decrease with $\eta$ as $1/\sqrt{\eta}$.
Some additional arguments in favor of this assumption
were presented in \cite{Q18}.
So, we suppose the following dependencies on $\eta$ for the parameters of the cluster correlation function:
\beq\label{Lam-fus}
\yetac=\yc/\sqrt{\eta}
 \ ,  \hs1
\Lametaz = const
\ .
\eeq
The values of the parameters
$\yc\!=\!2.7$ and $\Lametaz\!=\!0.8$
were chosen so that to obtain a correspondence with the values of the $\Sigma(n_F,\ n_B)$
obtained in \cite{Q18, EPJA19,UNIV19}.
Note that in these papers the $\Sigma(n_F,\ n_B)$ was calculated on the base of the string pair correlation function, $\Lambda(y_1-y_2)$,
extracted in \cite{NPA15} from the ALICE data \cite{ALICE15}
on the FB correlations
in the approximation of identical ("effective") strings,
which parameters were depended on initial energy in contrast with present consideration.

The $\SigFB$ calculated by formula (\ref{Sigfus}) with the listed parameters is presented in the figure \ref{Sig}
as function of the rapidity distance $\Dy$ between the centers of the observation windows
for two values of the window acceptance, $\dy$=0.2 and 0.4, and for two initial energies, 0.9 and 7 TeV.


It is instructive to compare the parameter dependencies on $\eta$ for string clusters, 
given by formulae (\ref{mean mult}) and (\ref{Lam-fus}):
\beq\label{cell-fus}
\mu_0^\eta=\mu_0^{1}\,\sqrt{\eta}
 \ ,  \hs1
\Lambda_0^{\eta} =\Lambda_0^{1}=const
 \ ,  \hs1
y_{corr}^{\eta}=\eta_{corr}^{1}/\sqrt{\eta}
\ ,
\eeq
with the ones
\beq\label{cell-nofus}
\mu_0^\eta=\mu_0^{1}\, \eta
 \ ,  \hs1
\Lambda_0^{\eta} =\Lambda_0^{1}/\eta
 \ ,  \hs1
y_{corr}^{\eta}=\eta_{corr}^{1}=const
\ .
\eeq
for the case, when we have not string fusion in the given transverse cell.
In the last case we find
$\SigFB =  \Sigma_1 (\muF, \muB)$, that 
in contrast with (\ref{Sigfus}) does not depends on $\alpha_\eta$.


\section{Conclusion}

In present research we performed the MC simulations
of string distributions
in the impact parameter plane
for pp interactions at few initial energies.
We took into account the string fusion processes leading to the formation of string clusters,
embedding
the finite lattice (the grid) in transverse plane.
Making the assumption on the properties of the string cluster correlation functions
we calculate the strongly intensive observable $\Sigma(\nF,\nB)$ for the minbias pp collisions
 at few initial energies and analyze its dependence on the distance between the centers
 of the observation windows and their width.

We show that the increase of the $\Sigma(\nF,\nB)$ with initial energy of pp collision
originates from
the
growth
of the portion of the fused string clusters
in arising string configurations.
In the case with different emitting clusters
this observable
becomes equal to the combination
of the ones for different clusters with
the weights depending
on details of the collision - its energy and centrality.
Analyzing
these dependencies of the $\Sigma(\nF,\nB)$
we can extract
the important information on the
characteristics
  of string clusters.

Similarly to the considered growth of the observable $\Sigma(\nF,\nB)$ with energy,
we expect also the enhance of the string fusion effects in the central pp collisions.
The analysis of the arising dependence of this observable
on the pp collision centrality
is now in progress (see preliminary results in \cite{Belokurova20}).	

\ack
The research was supported by the Russian Foundation for Basic Research grant 18-02-40075.

\end{document}